\documentclass[%
 reprint,
 showpacs,preprintnumbers,
 amsmath,amssymb,
 aps,
 prl,
]{revtex4-1}
\usepackage{dcolumn}
\usepackage[dvipdfm]{graphicx}
\usepackage{subfigure,color}
\usepackage{mathrsfs}

\makeatletter
\def\btt#1{\texttt{\@backslashchar#1}}
\DeclareRobustCommand\bblash{\btt{\@backslashchar}} \makeatother

\begin{document}

\title{Three-dimensional Symmetry-breaking Topological Matters}

\author{Tetsuro Habe$^1$ and Yasuhiro Asano$^{1,2}$}
\affiliation{$^1$Department of Applied Physics,
Hokkaido University, Sapporo 060-8628, Japan}
\affiliation{$^2$Center for Topological Science \& Technology,
Hokkaido University, Sapporo 060-8628, Japan}

\date{\today}

\begin{abstract}
We discuss topological electronic states described by the Dirac
Hamiltonian plus an additional one
in three-dimension. When the additional Hamiltonian is an element of an
Abelian group,
electronic states become topologically nontrivial even in the absence of
the fundamental symmetries
such as the time-reversal symmetry and the particle-hole one.
Such symmetry-breaking topological states are characterized by the Chern
number defined in the two-dimensional partial Brillouin zone. 
The topological insulators in Zeeman fields
are an example of
the symmetry-breaking topological matters. We show the crossover from
the topological insulating
phase to the topological semimetal one in strong Zeeman fields.
\end{abstract}

\pacs{73.20.At, 73.20.Hb}

\maketitle

Topological classification has successfully predicted a number of topologically nontrivial electronic states in the condensed matters.
 Each topological phase is characterized by a topological number defined in the presence 
of the fundamental symmetries preserved in the materials such as the time-reversal symmetry in topological insulators\cite{Fu2007,Fu2007-2,Moore2007,Chen2009}, 
the particle-hole symmetry in superfluids and topological superconductors\cite{Qi2009,Fu2010TS}, 
and the crystal symmetry in topological crystalline insulators\cite{Fu2011,Hsieh2012}.
The table of the topological classes~\cite{Schnyder2008,Chiu2013} has suggested a close relationship between the 
appearance of the topological phase and the invariance of the Hamiltonian under the fundamental symmetries.
However, the surface states, an evidence of the topological phase, often remain gapless even when the fundamental symmetries 
are broken by perturbations. For instance, $Z_2$ number in the presence of the time-reversal symmetry characterizes the 
topological state of $^3$He-B phase belonging to class DIII. In the Zeeman field, however, $^3$He-B phase still hosts the gapless states
on a surface parallel to the Zeeman field. 
Mizushima et. al.~\cite{Mizushima2012} have explained the existence of the gapless surface 
states in terms of a topological number defined by using the remaining symmetry of $^3$He-B phase under the Zeeman field. 
Finding a particular topological number for explaining a particular gapless surface states is a way to understand physics of 
topological materials. However there may be more general route to characterize topological electronic states. 

In this paper, we will show an alternative way to search the topological materials in three-dimension
in the absence of the fundamental symmetries.
The electronic structures of topological materials are described by the $4\times 4$ Dirac Hamiltonian. 
A unitary transformation in the two-dimensional partial Brillouin zone (BZ) deforms the $4\times 4$ Dirac Hamiltonian into 
two decoupled $2\times 2$ quantum Hall Hamiltonians whose electric states are characterized by the nontrivial Chern number. 
The Chern number characterizes the two-dimensional topological states in class A where the Hamiltonian 
are not necessary to preserve any fundamental symmetries. 
Therefore the Chern number in each quantum Hall state (QHS) remains unchanged as 
far as perturbations do not mix them. 
Such perturbed Hamiltonian preserves a certain symmetry in a two-dimensional BZ
and forms an Abelian group.
Generally speaking, the symmetry in a two-dimensional partial BZ are not necessary to be 
held in whole BZ in three-dimension.
Therefore, the Dirac Hamiltonian plus such perturbed Hamiltonian describe topological states 
characterized by the Chern number without any fundamental symmetries. 
Our theory provides more general framework for searching topologically nontrivial states than the table of the topological 
classes~\cite{Schnyder2008,Chiu2013}.
On the basis of the prescription, we propose semimetal phases of time-reversal invariant topological insulator 
under strong Zeeman field. Such semimetals are an example of the symmetry-breaking topological materials because 
they are characterized only by the Chern number in the partial BZ.

We begin with the three-dimensional Dirac Hamiltonian which describes electronic structures 
of topological materials such as topological insulators, 
topological superconductors, and 
superfluids,
\begin{align}
H_0=&a\alpha^{\mu}p_{\mu}+M\beta,\;\;\;\mu=x,y,z \label{Hamiltonian},\\
M=&(m-b\boldsymbol{p}^2),
\end{align}
where $a$, $b$, and $m$ are positive constants, $\alpha$ and $\beta$ are the $4\times4$ Dirac matrices,
\begin{align*}
\alpha^{\mu}=\sigma^\mu\tau^x
=\begin{pmatrix}
0&\sigma^{\mu}\\
\sigma^{\mu}&0
\end{pmatrix}
,\;\;\;
\beta=\sigma^0\tau^z
\nonumber.
\end{align*}
The diagonal $2\times2$ blocked sectors describe the two orbital spaces in the topological insulators\cite{Liu2010} or the particle-hole subspace in the 
topological superconductors.
Here the Pauli matrices $\sigma^\mu$ and $\tau^\mu$ for $\mu=x,y,z$ act on the spin and the orbital indexces of an electron, respectively.
The matrices $\sigma^0$ and $\tau^0$ are the $2\times2$ identity matrix in the corresponding subspace.
The index which appears twice in a single term means the summation for $\mu=x,y,z$.

\begin{figure}[htbp]
 \includegraphics[width=85mm]{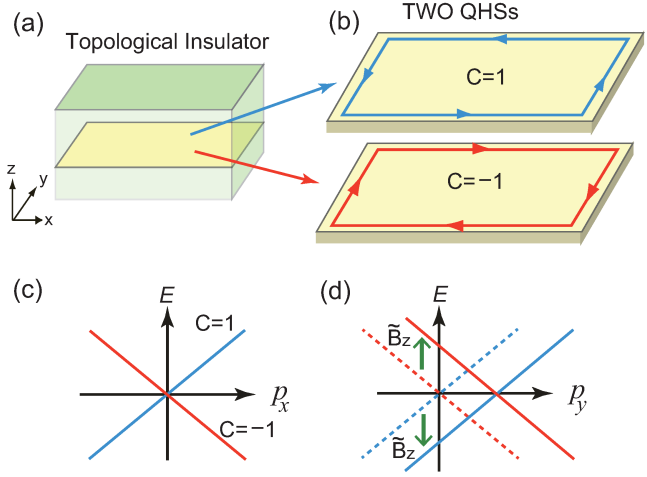}
\caption{In (a), a plane represents the two-dimensional partial Brillouin zone (BZ) embedded in the three-dimensional one.
In the partial BZ, the two blocked Hamiltonian $h(M)$ and $h(-M)$ describe the two quantum Hall states (QHSs)
with the opposite chiral edge mode as shown in (b), where
 the arrows denote the direction of the chiral edge current.
The corresponding dispersion of the edge modes is shown in (c).
Fig.~(d) illustrates the dispersion of the edge modes under the antisymmetric Zeeman field which 
shifts the chemical potential of the two QHSs inversely.
 As a consequence, the Dirac point moves to another point in the BZ. 
 }\label{fig1}
\end{figure}

We first focus on a partial Brillouin zone specified by $p_z=0$.
The Hamiltonian can be separated into two blocks as
\begin{align}
H^{U}_0=&\begin{pmatrix}
\boldsymbol{d}(M)\cdot\boldsymbol{\sigma}&0\\
0&\boldsymbol{d}(-M)\cdot\boldsymbol{\sigma}
\end{pmatrix},\label{Qimodel}\\
\boldsymbol{d}(M)=&(ap_x,ap_y,M_{p_z=0}),\label{dm}
\end{align}
by applying the unitary transformation $H^{U}=U HU^\dagger$ with
\begin{align}
U=\begin{pmatrix}
1&0&0&0\\
0&0&0&1\\
0&0&1&0\\
0&1&0&0
\end{pmatrix}\label{U}.
\end{align}
The Hamiltonian of the blocked sector $h(\pm M)=\boldsymbol{d}(\pm M)\cdot\boldsymbol{\sigma}$ is a simple model of 
the Quantum Hall systems~\cite{Qi2008} as shown in Fig.~\ref{fig1}(b).
The chiral edge modes of the two quantum Hall states (QHSs) have the opposite chirality to each other.
When the Chern number C is 1 for $h(M)$, it is -1 for $h(-M)$.
The dispersion of the chiral edge modes
goes across the zero-energy at so called Dirac point in $p_x$ and $p_y$ axes on the $xz$ and $yz$ surfaces, respectively as 
schematically shown in Fig.~\ref{fig1}(c).
The Dirac point is stable even when we add a perturbation $H_{\mathcal{P}}^U$ to $H^U_0$ 
as far as $H_{\mathcal{P}}^U$ is invariant under the unitary transformation given
by $\mathcal{P}^U_{xy}=\tau^z$ (i.e., $\mathcal{P}^U_{xy}H_{\mathcal{P}}^U (\mathcal{P}^U_{xy})^\dagger=H_{\mathcal{P}}^U $). 
This is because such $H_{\mathcal{P}}^U$ does not mix the two QHSs.
In the original representation before applying $U$, 
the perturbed Hamiltonian $H_{\mathcal{P}}$ should satisfy $\mathcal{P}_{xy} H_{\mathcal{P}} \mathcal{P}_{xy}^\dagger = H_{\mathcal{P}}$, where
 $\mathcal{P}_{xy}=\sigma^z\tau^z$ represents the combined operation of the pseudo spin inversion 
 $\tau^\mu\to -\tau^\mu$ and spin inversion $\sigma^\mu\to -\sigma^\mu$ for $\mu=x$ and $y$.
The set of Hermite matrices $H_{\mathcal{P}}$ forms an Abelian group $A_{P}$ whose elements are invariant under the unitary transformation by 
$\mathcal{P}_{xy}$. 
Non-trivial Chern numbers in the partial BZ of $p_xp_y$ plane
explain the existence of the gapless states on the four surfaces perpendicular to the $xy$ plane. 
It is possible to generalize the argument above to the partial BZ specified by $p_\rho=0$. 
In this case, the unitary operator is represented by
\begin{align}
\mathcal{P}_{\mu\nu}=\epsilon_{\mu\nu\rho}\sigma^\rho\tau^z,\label{C1}
\end{align}
where $\epsilon_{\mu\nu\rho}$ is antisymmetric symbol.
Here we supply an example of $H_{\mathcal{P}}$, 
\begin{align}
\left(
\begin{array}{cc}
a_0 + a_\rho \sigma^\rho  & b_\mu \sigma^\mu + b_\nu \sigma^\nu \\
b_\mu^\ast \sigma^\mu + b_\nu^\ast \sigma^\nu & c_0 + c_\rho \sigma^\rho
\end{array}\right), \label{rhp}
\end{align}
where $\mu$, $\nu$ and $\rho$ represent $x$, $y$ and $z$, and they are not equal to one another. 
Coefficients $a_0$, $a_\rho$, $c_0$ and $c_\rho$ are real numbers but they are not necessary to be constants. 

The Abelian group $A_P$ has a subgroup of $A_{PQ}$. 
The perturbed Hamiltonian belonging to $A_{PQ}$ does not remove 
the gapless states on all the surfaces. To show this property, we 
consider the elements commute with $\mathcal{P}_{yz}$. The gapless states remain on the four surfaces 
perpendicular to the $yz$ plane 
under such perturbed Hamiltonian. Thus we show the gapless states on the remaining two surfaces parallel 
to the $yz$ plane.
After applying the transformation by $U$ in Eq.~(\ref{U}), Eq.~(\ref{Hamiltonian}) with $p_y=p_z=0$ becomes 
\begin{align}
H_0^{WU}= W H^U _0 W^\dagger=&\begin{pmatrix}
\boldsymbol{d}'(M)\cdot\boldsymbol{\sigma}&0\\
0&\boldsymbol{d}'(M)\cdot\boldsymbol{\sigma}
\end{pmatrix},\label{Gamma}\\
\boldsymbol{d}'(M)=&(ap_x,0,m-bp_x^2)\nonumber,
\end{align}
where $W$ is a unitary matrix of $W=\mathrm{diag}[\sigma^0,-i\sigma^x]$.
At the point of $p_y=p_z=0$, the edge states on the $yz$ surface are degenerate 
at the zero-energy. 
This is because 
the two blocked sectors are equivalent to each other in this representation
and because each blocked sector preserves the 'particle-hole symmetry' within the $2\times 2$ space represented by a relation 
$\sigma^y(\boldsymbol{d}'\cdot\boldsymbol{\sigma})\sigma^y=-\boldsymbol{d}'\cdot\boldsymbol{\sigma}$.
Thus the degeneracy of the two edge states remains even when 
we add the 'particle-hole' symmetrical Hamiltonian $H^{WU}_{\mathcal{Q}}$ to $H_0^{WU}$.
Such Hamiltonian $H^{WU}_{\mathcal{Q}}$ should satisfy the relation 
$\mathcal{Q}^{WU}_{yz} H^{WU}_{\mathcal{Q}} (\mathcal{Q}^{WU}_{yz})^\dagger= -H^{WU}_{\mathcal{Q}}$
with $ \mathcal{Q}^{WU}_{yz} = \sigma^y\tau^0$.
The Hermitian matrices form the Abelian subgroup $A_{PQ}$
whose elements $H_{\mathcal{PQ}}$ satisfy $\mathcal{Q}_{yz} H_{\mathcal{PQ}} \mathcal{Q}_{yz}^\dagger =-H_{\mathcal{PQ}}$ 
and $\mathcal{P}_{yz} H_{\mathcal{PQ}} \mathcal{P}_{yz}^\dagger = H_{\mathcal{PQ}}$ at the same time.
In the original representation before applying the transformation of $W$ and $U$,  
the unitary matrix $\mathcal{Q}_{yz}$ is given by
\begin{align}
\mathcal{Q}_{yz}=\sigma^x\tau^y.\label{q1}
\end{align}
When we consider elements invariant under $\mathcal{P}_{\mu\nu}$ in general, 
the unitary matrix in Eq.~(\ref{q1}) is represented as $\mathcal{Q}_{\mu\nu}=\epsilon_{\mu\nu\rho}\sigma^\rho \tau^y$.
In addition to the elements in $A_{PQ}$, the Hamiltonian proportional to 
$4 \times 4$ identity matrix does not affect the gapless surface states at all.

In a short summary, we have discussed
the three-dimensional symmetry-breaking topological 
matter whose electronic structures are 
represented by
\begin{align}
H=H_0+H_{\mathcal{P}},
\end{align}
where $H_{\mathcal{P}}$ commutes with $\mathcal{P}_{\mu\nu}=\epsilon_{\mu\nu\rho}\sigma^\rho\tau^z$. 
The set of such Hamiltonian forms the Abelian group $A_{\mathcal{P}}$.  
The Chern number defined in the two-dimensional partial BZ specified by $p_\rho=0$ characterizes 
the topologically nontrivial states.
Generally speaking, the symmetry in a two-dimensional partial BZ are not necessary to be 
held in whole BZ in three-dimension. In addition, the gapped energy spectra in a two-dimensional partial BZ
are not necessary to be held in whole BZ in three-dimension.
The word "symmetry breaking" means that the symmetries preserved in $H_0$ are broken by $H_{\mathcal{P}}$, 
and those preserved in $H_{\mathcal{P}}$ are broken by $H_0$.
In the following, we discuss electronic states of three-dimensional topological insulator under the Zeeman field
as an example of the three-dimensional symmetry-breaking topological matter. 


The Zeeman field is represented by two vectors $\boldsymbol{B}$ and $\tilde{\boldsymbol{B}}$ as
\begin{align}
H_1=&B_{\mu}\sigma^\mu\tau^0+\tilde{B}_\mu\sigma^\mu\tau^z,\label{Zeeman}
\end{align}
where $B_{\mu}\sigma^\mu\tau^0$ and $\tilde{B}_\mu\sigma^\mu\tau^z$ are the symmetric and the antisymmetric parts of the Zeeman field 
with respect to the two orbitals, respectively.
The antisymmetric part of $\tilde{B}_\mu\sigma^\mu\tau^z$ is attributed to the difference of the coupling constants 
to the Zeeman field in the two orbitals and is dominant in Bi$_2$Se$_3$\cite{Liu2010}.
We consider the subgap states on the $yz$ surface for a while.
When the weak Zeeman field of $B, \tilde{B}<m$ is applied along the $z$ axis parallel to the $yz$ surface, 
the two QHSs in the two-dimensional partial BZ on $p_yp_z$ plane
still remain decoupled from each other
because the $H_1$ commute with $\mathcal{P}_{xy}$. 
Applying the unitary transformation of $U$, the Hamiltonian becomes
\begin{align}
H^U_1=B_z\sigma^z\tau^0
+
\tilde{B}_z\sigma^0\tau^z. \label{z-direction}
\end{align}
The symmetric Zeeman field $B_z$ gives a constant correction to $M$ in Eq.~(\ref{Qimodel}) and 
does not affect the Dirac point in two-dimensional BZ at all.
On the other hand, the antisymmetric Zeeman field $\tilde{B}_z$  
shifts the chemical potential of the two QHSs inversely. As a result, the Dirac point 
moves from the $\Gamma$-point $(p_y,p_z)=0$ as shown in Fig.~\ref{fig1}(d).
The situation is similar to the shift of the Dirac point at the interface facing to a ferromagnetic insulator\cite{Tanaka2009,Liu2009,Wray2010,Habe2012}.

Next we consider the Zeeman field in the direction perpendicular to the $yz$ surface.
We conclude that the symmetric Zeeman field 
would remove the gapless states from the $yz$ surface 
because it does not belong to the Abelian subgroup $A_{PQ}$.
On the other hand, the antisymmetric Zeeman field leaves the gapless states because it
belongs to the Abelian subgroup $A_{PQ}$. It is easy to confirm these conclusions by the argument below.
Both the symmetric $B_x \sigma^x \tau^0$ and the antisymmetric $\tilde{B}_x \sigma^x \tau^z$ Zeeman field 
commute with $\mathcal{P}_{yz}=\sigma^x\tau^z$ but do not commute with either $\mathcal{P}_{xy}=\sigma^z\tau^z$ or 
$\mathcal{P}_{xz}=-\sigma^y\tau^z$. 
The symmetric Zeeman field does not anticommute with $Q_{yz}$ in Eq.~(\ref{q1}), which indicates the gapless 
states are no longer guaranteed on $yz$ surface.
On the other hand, the antisymmetric Zeeman field anticommutes with $Q_{yz}$, which means the antisymmetric
Zeeman field belongs to $A_{PQ}$. 
This is because the antisymmetric Zeeman field preserves the 'particle-hole symmetry' in each QHS.
As a result, the symmetric (antisymmetric) Zeeman field removes (leaves) the gapless energy 
spectra on the surface perpendicular to the Zeeman field.
So far we have considered the weak Zeeman field of $m>|B|$ and $m>|\tilde{B}|$.
In the strong Zeeman field $m<|B|$ and $m<|\tilde{B}|$, another topological phases
would be also expected because the two QHSs still remain decoupled from each other.
We will confirm the last statement by numerical calculation. 

In the following, we confirm the analysis above by the numerical calculation on the tight-binding model,
\begin{align}
H=&\sum_{\boldsymbol{p}}\boldsymbol{c}({\boldsymbol{p}})^\dagger \left[(m-b\sum_{\nu=x,y,z}(1-\cos p_\nu))\sigma^0\tau^z\right.\nonumber\\
&\left.+a\sin p_\mu\sigma^\mu\tau^x+\sigma^\nu(B_\nu\tau^0+\tilde{B}_\nu\tau^z)\right]\boldsymbol{c}({\boldsymbol{p}})
\end{align}
with the parameters $m$, $a$ and $b$ used in Ref.~\onlinecite{Habe2012}.
Here $\boldsymbol{c}(\boldsymbol{p})=(c_{1\uparrow},c_{1\downarrow},c_{2\uparrow},c_{2\downarrow})^T$ is the annihilation operator with four 
components corresponding spin $\uparrow, \downarrow$ and orbital $1, 2$ subspaces.
To calculate the energy spectra of the surface state on $yz$ plane, 
we consider the lattice along $x$ axis with
\begin{align}
\sum_{p_x}\cos p_x c^\dagger(p_x) c(p_x)\rightarrow \frac{1}{2}\sum_{j}(c^\dagger(j+1) c(j)+h.c.),\nonumber\\
\sum_{p_x}\sin p_x c^\dagger(p_x) c(p_x)\rightarrow \frac{1}{2i}\sum_{j}(c^\dagger(j+1) c(j)-h.c.),\nonumber
\end{align}
where we utilize $j$ as the position on the $x$ axis and employee the hard-wall boundary condition in the $x$ axis.
\begin{figure}[htbp]
 \includegraphics[width=75mm]{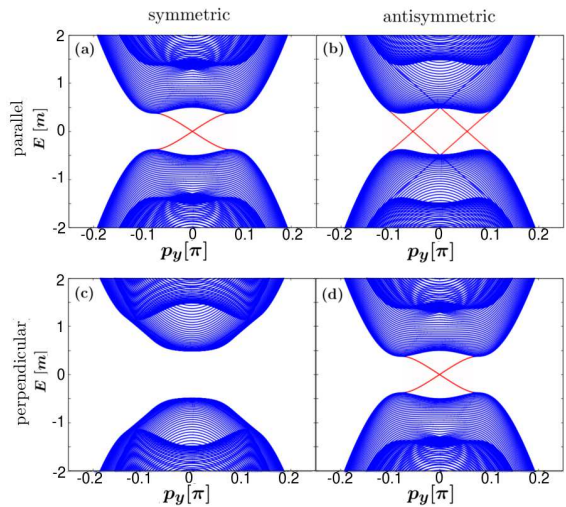}
 \caption{The energy spectra of the $yz$ surface states under the symmetric Zeeman field (a) and (c) or the antisymmetric Zeeman field (b) and (d). 
The Zeeman field is applied in a parallel direction to the surface in (a) and (b).
In (c) and (d), the Zeeman field is applied in the perpendicular direction to the surface. 
The momentum in the $z$ direction is fixed at $p_z=0$ in all figures. 
 }\label{fig4}
\end{figure}

At first, we consider the weak Zeeman field. 
The topological phases can be confirmed by the appearance of the gapless surface states. 
 In Fig.~\ref{fig4}, we show the energy spectra of the surface states on $yz$ plane under the 
 symmetric Zeeman field with $|B|=m/2$ and $|\tilde{B}|=0$ in (a) and (c),
 and those under the antisymmetric Zeeman field $|B|=0$ and $|\tilde{B}|=m/2$ in (b) and (d).
The results for the Zeeman field parallel to the $yz$ surface are shown in Fig.~\ref{fig4}(a) and (b).
The energy spectra on the surface remain gapless in both the symmetric and the antisymmetric Zeeman field.
Especially, under the antisymmetric field in Fig.~\ref{fig4}(b), the Dirac point shifts from the $\Gamma$ point.
When the symmetric Zeeman field is in the perpendicular to the surface, 
the gapless states on $yz$ plane vanish as shown in Fig.~\ref{fig4}(c).
On the other hand, as shown in Fig.~\ref{fig4}(d), the gapless states survive 
under the antisymmetric Zeeman field. The numerical results in Fig.~\ref{fig4} suggest the validity of the analysis.

\begin{figure}[htbp]
 \includegraphics[width=80mm]{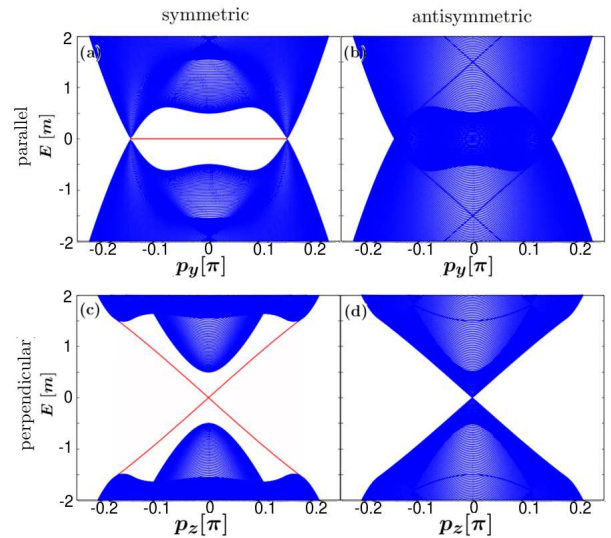}
 \caption{The energy spectra on the $yz$ surface under the large Zeeman field applied in the $y$ direction.
The results are plotted along the $p_y$ axis with $p_z=0$ (upper figures) and along the $p_z$ axis with $p_y=0$ (lower figures). 
The left figures of (a) and (c) are the results for the symmetric Zeeman field at $B_y=1.5m$. 
The right figures of (b) and (d) are calculated under the antisymmetric Zeeman field at $\tilde{B}_y=1.5m$. 
 }\label{fig7}
\end{figure}

Finally, we show the numerical results for the large Zeeman field $m<|B|$ and $m<|\tilde{B}|$.
According to Eq.~(\ref{Qimodel}) and (\ref{z-direction}), the large enough symmetric Zeeman field 
equalizes two the Chern numbers of the two QHSs because of $\mathrm{sgn}[M+B]=\mathrm{sgn}[-M+B]$.
Therefore, the net Chern number becomes nontrivial in the two-dimensional BZ.
The electronic states of the topological insulator under the large symmetric Zeeman field 
consist of a number of QHSs with the same Chern number stacking in momentum space.
As a consequence, electronic states become the Weyl semimetal phase~\cite{Murakami2007,Wan2011,Yang2011,Burkov2011}. 
In Fig.~\ref{fig7}(a) and (c), we show the energy spectra on $yz$ surface under the strong symmetric Zeeman field in the 
$y$ axis with $B_y=1.5 m$. 
The semimetal hosts the chiral surface modes so called the Fermi arc
as shown by
a pair of linear dispersion in (c). The spectra in Fig.~\ref{fig7} include the contribution from the 
two $yz$ planes parallel to the Zeeman field. 
When one $yz$ surface hosts a chiral modes with the positive velocity, the other $yz$ surface hosts a chiral modes with 
the negative velocity.
The results are consistent with a study of the large Zeeman field in Ref.~\onlinecite{Liu2013}.
On the other hand, the antisymmetric Zeeman field biases the two QHSs reversely but holds the Chern number unchanged.
As a result, the topological insulator qualitatively changes into the nodal semimetal 
as shown in Fig.~\ref{fig7}(b) and (d), where we show the excitation spectra of bulk states under the 
strong antisymmetric Zeeman field in the $y$ axis with $\tilde{B}_y=1.5 m$.
The numerical results in Fig.~\ref{fig7} might imply the applicability of our approach to metallic materials.
 
In summary, we have discussed topologically nontrivial electronic states described by the Hamiltonian of $H=H_0+H_{\mathcal{P}}$, 
where $H_0$ is the Dirac Hamiltonian, 
and $H_{\mathcal{P}}$ is an additional Hamiltonian belonging to an Abelian group of $A_{P}$.
The Chern number defined in the two-dimensional partial Brillouin zone characterizes the topological states. 
Generally speaking, such topological states can be realized without any fundamental symmetries such as 
the time-reversal, the particle-hole and the chiral symmetries.
Thus we have proposed a possibility of symmetry-breaking topological matters in three-dimension.
When $H_{\mathcal{P}}$ belongs to an Abelian subgroup of $A_{PQ}$, the symmetry-breaking topological matter has the gapless states on all the surfaces.
We have numerically studied the electronic states in a topological insulator in Zeeman fields
as an example of the three-dimensional symmetry-breaking topological matter. 
The results of the low-energy excitation spectra suggest the Weyl semimetal phase or the line-nodal semimetal one 
depending on characters of large Zeeman fields.

The authors are grateful to K. Nomura for useful comments on our manuscript.
This work was supported by 
the "Topological Quantum Phenomena" (No. 22103002) Grant-in Aid for 
Scientific Research on Innovative Areas from the Ministry of Education, 
Culture, Sports, Science and Technology (MEXT) of Japan.
\bibliography{TI}

\providecommand{\noopsort}[1]{}\providecommand{\singleletter}[1]{#1}%
\begin{thebibliography}{22}%
\makeatletter
\providecommand \@ifxundefined [1]{%
 \@ifx{#1\undefined}
}%
\providecommand \@ifnum [1]{%
 \ifnum #1\expandafter \@firstoftwo
 \else \expandafter \@secondoftwo
 \fi
}%
\providecommand \@ifx [1]{%
 \ifx #1\expandafter \@firstoftwo
 \else \expandafter \@secondoftwo
 \fi
}%
\providecommand \natexlab [1]{#1}%
\providecommand \enquote  [1]{``#1''}%
\providecommand \bibnamefont  [1]{#1}%
\providecommand \bibfnamefont [1]{#1}%
\providecommand \citenamefont [1]{#1}%
\providecommand \href@noop [0]{\@secondoftwo}%
\providecommand \href [0]{\begingroup \@sanitize@url \@href}%
\providecommand \@href[1]{\@@startlink{#1}\@@href}%
\providecommand \@@href[1]{\endgroup#1\@@endlink}%
\providecommand \@sanitize@url [0]{\catcode `\\12\catcode `\$12\catcode
  `\&12\catcode `\#12\catcode `\^12\catcode `\_12\catcode `\%12\relax}%
\providecommand \@@startlink[1]{}%
\providecommand \@@endlink[0]{}%
\providecommand \url  [0]{\begingroup\@sanitize@url \@url }%
\providecommand \@url [1]{\endgroup\@href {#1}{\urlprefix }}%
\providecommand \urlprefix  [0]{URL }%
\providecommand \Eprint [0]{\href }%
\providecommand \doibase [0]{http://dx.doi.org/}%
\providecommand \selectlanguage [0]{\@gobble}%
\providecommand \bibinfo  [0]{\@secondoftwo}%
\providecommand \bibfield  [0]{\@secondoftwo}%
\providecommand \translation [1]{[#1]}%
\providecommand \BibitemOpen [0]{}%
\providecommand \bibitemStop [0]{}%
\providecommand \bibitemNoStop [0]{.\EOS\space}%
\providecommand \EOS [0]{\spacefactor3000\relax}%
\providecommand \BibitemShut  [1]{\csname bibitem#1\endcsname}%
\let\auto@bib@innerbib\@empty
\bibitem [{\citenamefont {Fu}\ \emph {et~al.}(2007)\citenamefont {Fu},
  \citenamefont {Kane},\ and\ \citenamefont {Mele}}]{Fu2007}%
  \BibitemOpen
  \bibfield  {author} {\bibinfo {author} {\bibfnamefont {L.}~\bibnamefont
  {Fu}}, \bibinfo {author} {\bibfnamefont {C.~L.}\ \bibnamefont {Kane}}, \ and\
  \bibinfo {author} {\bibfnamefont {E.~J.}\ \bibnamefont {Mele}},\ }\href@noop
  {} {\bibfield  {journal} {\bibinfo  {journal} {Phys.\ Rev. \ Lett.}\ }\textbf
  {\bibinfo {volume} {98}},\ \bibinfo {pages} {106803} (\bibinfo {year}
  {2007})}\BibitemShut {NoStop}%
\bibitem [{\citenamefont {Fu}\ and\ \citenamefont {Kane}(2007)}]{Fu2007-2}%
  \BibitemOpen
  \bibfield  {author} {\bibinfo {author} {\bibfnamefont {L.}~\bibnamefont
  {Fu}}\ and\ \bibinfo {author} {\bibfnamefont {C.~L.}\ \bibnamefont {Kane}},\
  }\href@noop {} {\bibfield  {journal} {\bibinfo  {journal} {Phys.\ Rev. \ B.}\
  }\textbf {\bibinfo {volume} {76}},\ \bibinfo {pages} {045302} (\bibinfo
  {year} {2007})}\BibitemShut {NoStop}%
\bibitem [{\citenamefont {Moore}\ and\ \citenamefont
  {Balents}(2007)}]{Moore2007}%
  \BibitemOpen
  \bibfield  {author} {\bibinfo {author} {\bibfnamefont {J.~E.}\ \bibnamefont
  {Moore}}\ and\ \bibinfo {author} {\bibfnamefont {L.}~\bibnamefont
  {Balents}},\ }\href@noop {} {\bibfield  {journal} {\bibinfo  {journal}
  {Phys.\ Rev. \ B.}\ }\textbf {\bibinfo {volume} {75}},\ \bibinfo {pages}
  {121306} (\bibinfo {year} {2007})}\BibitemShut {NoStop}%
\bibitem [{\citenamefont {Chen}\ \emph {et~al.}(2009)\citenamefont {Chen},
  \citenamefont {Analytis}, \citenamefont {Chu}, \citenamefont {Liu},
  \citenamefont {Mo}, \citenamefont {Qi}, \citenamefont {Zhang}, \citenamefont
  {Lu}, \citenamefont {Dai}, \citenamefont {Fang}, \citenamefont {Zhang},
  \citenamefont {Fisher}, \citenamefont {Hussain},\ and\ \citenamefont
  {Shen}}]{Chen2009}%
  \BibitemOpen
  \bibfield  {author} {\bibinfo {author} {\bibfnamefont {Y.~L.}\ \bibnamefont
  {Chen}}, \bibinfo {author} {\bibfnamefont {J.~G.}\ \bibnamefont {Analytis}},
  \bibinfo {author} {\bibfnamefont {J.-H.}\ \bibnamefont {Chu}}, \bibinfo
  {author} {\bibfnamefont {Z.~K.}\ \bibnamefont {Liu}}, \bibinfo {author}
  {\bibfnamefont {S.~K.}\ \bibnamefont {Mo}}, \bibinfo {author} {\bibfnamefont
  {X.~L.}\ \bibnamefont {Qi}}, \bibinfo {author} {\bibfnamefont {H.~J.}\
  \bibnamefont {Zhang}}, \bibinfo {author} {\bibfnamefont {D.~H.}\ \bibnamefont
  {Lu}}, \bibinfo {author} {\bibfnamefont {X.}~\bibnamefont {Dai}}, \bibinfo
  {author} {\bibfnamefont {Z.}~\bibnamefont {Fang}}, \bibinfo {author}
  {\bibfnamefont {S.~C.}\ \bibnamefont {Zhang}}, \bibinfo {author}
  {\bibfnamefont {I.~R.}\ \bibnamefont {Fisher}}, \bibinfo {author}
  {\bibfnamefont {Z.}~\bibnamefont {Hussain}}, \ and\ \bibinfo {author}
  {\bibfnamefont {Z.~X.}\ \bibnamefont {Shen}},\ }\href@noop {} {\bibfield
  {journal} {\bibinfo  {journal} {Science}\ }\textbf {\bibinfo {volume}
  {325}},\ \bibinfo {pages} {178} (\bibinfo {year} {2009})}\BibitemShut
  {NoStop}%
\bibitem [{\citenamefont {Qi}\ \emph {et~al.}(2009)\citenamefont {Qi},
  \citenamefont {Hughes}, \citenamefont {Raghu},\ and\ \citenamefont
  {Zhang}}]{Qi2009}%
  \BibitemOpen
  \bibfield  {author} {\bibinfo {author} {\bibfnamefont {X.-L.}\ \bibnamefont
  {Qi}}, \bibinfo {author} {\bibfnamefont {T.~L.}\ \bibnamefont {Hughes}},
  \bibinfo {author} {\bibfnamefont {S.}~\bibnamefont {Raghu}}, \ and\ \bibinfo
  {author} {\bibfnamefont {S.-C.}\ \bibnamefont {Zhang}},\ }\href@noop {}
  {\bibfield  {journal} {\bibinfo  {journal} {Phys. Rev. Lett.}\ }\textbf
  {\bibinfo {volume} {102}},\ \bibinfo {pages} {187001} (\bibinfo {year}
  {2009})}\BibitemShut {NoStop}%
\bibitem [{\citenamefont {Fu}\ and\ \citenamefont {Berg}(2010)}]{Fu2010TS}%
  \BibitemOpen
  \bibfield  {author} {\bibinfo {author} {\bibfnamefont {L.}~\bibnamefont
  {Fu}}\ and\ \bibinfo {author} {\bibfnamefont {E.}~\bibnamefont {Berg}},\
  }\href@noop {} {\bibfield  {journal} {\bibinfo  {journal} {Phys. Rev. Lett.}\
  }\textbf {\bibinfo {volume} {105}},\ \bibinfo {pages} {097001} (\bibinfo
  {year} {2010})}\BibitemShut {NoStop}%
\bibitem [{\citenamefont {Fu}(2011)}]{Fu2011}%
  \BibitemOpen
  \bibfield  {author} {\bibinfo {author} {\bibfnamefont {L.}~\bibnamefont
  {Fu}},\ }\href@noop {} {\bibfield  {journal} {\bibinfo  {journal} {Phys.\
  Rev.\ Lett.}\ }\textbf {\bibinfo {volume} {106}},\ \bibinfo {pages} {106802}
  (\bibinfo {year} {2011})}\BibitemShut {NoStop}%
\bibitem [{\citenamefont {Hsieh}\ \emph {et~al.}(2012)\citenamefont {Hsieh},
  \citenamefont {Lin}, \citenamefont {Liu}, \citenamefont {Duan}, \citenamefont
  {Bansil},\ and\ \citenamefont {Fu}}]{Hsieh2012}%
  \BibitemOpen
  \bibfield  {author} {\bibinfo {author} {\bibfnamefont {T.~H.}\ \bibnamefont
  {Hsieh}}, \bibinfo {author} {\bibfnamefont {H.}~\bibnamefont {Lin}}, \bibinfo
  {author} {\bibfnamefont {J.}~\bibnamefont {Liu}}, \bibinfo {author}
  {\bibfnamefont {W.}~\bibnamefont {Duan}}, \bibinfo {author} {\bibfnamefont
  {A.}~\bibnamefont {Bansil}}, \ and\ \bibinfo {author} {\bibfnamefont
  {L.}~\bibnamefont {Fu}},\ }\href@noop {} {\bibfield  {journal} {\bibinfo
  {journal} {Nat. Comm.}\ }\textbf {\bibinfo {volume} {3}},\ \bibinfo {pages}
  {982} (\bibinfo {year} {2012})}\BibitemShut {NoStop}%
\bibitem [{\citenamefont {Schnyder}\ \emph {et~al.}(2008)\citenamefont
  {Schnyder}, \citenamefont {Ryu}, \citenamefont {Furusaki},\ and\
  \citenamefont {Ludwig}}]{Schnyder2008}%
  \BibitemOpen
  \bibfield  {author} {\bibinfo {author} {\bibfnamefont {A.~P.}\ \bibnamefont
  {Schnyder}}, \bibinfo {author} {\bibfnamefont {S.}~\bibnamefont {Ryu}},
  \bibinfo {author} {\bibfnamefont {A.}~\bibnamefont {Furusaki}}, \ and\
  \bibinfo {author} {\bibfnamefont {A.~W.~W.}\ \bibnamefont {Ludwig}},\
  }\href@noop {} {\bibfield  {journal} {\bibinfo  {journal} {Phys. \ Rev. \ B}\
  }\textbf {\bibinfo {volume} {78}},\ \bibinfo {pages} {195125} (\bibinfo
  {year} {2008})}\BibitemShut {NoStop}%
\bibitem [{\citenamefont {Chiu}\ \emph {et~al.}(shed)\citenamefont {Chiu},
  \citenamefont {Yao},\ and\ \citenamefont {Ryu}}]{Chiu2013}%
  \BibitemOpen
  \bibfield  {author} {\bibinfo {author} {\bibfnamefont {C.-K.}\ \bibnamefont
  {Chiu}}, \bibinfo {author} {\bibfnamefont {H.}~\bibnamefont {Yao}}, \ and\
  \bibinfo {author} {\bibfnamefont {S.}~\bibnamefont {Ryu}},\ }\href@noop {}
  {\bibfield  {journal} {\bibinfo  {journal} {arXiv}\ ,\ \bibinfo {pages}
  {1303.1843}} (\bibinfo {year} {unpublished})}\BibitemShut {NoStop}%
\bibitem [{\citenamefont {Mizushima}\ \emph {et~al.}(2012)\citenamefont
  {Mizushima}, \citenamefont {Sato},\ and\ \citenamefont
  {Machida}}]{Mizushima2012}%
  \BibitemOpen
  \bibfield  {author} {\bibinfo {author} {\bibfnamefont {T.}~\bibnamefont
  {Mizushima}}, \bibinfo {author} {\bibfnamefont {M.}~\bibnamefont {Sato}}, \
  and\ \bibinfo {author} {\bibfnamefont {K.}~\bibnamefont {Machida}},\
  }\href@noop {} {\bibfield  {journal} {\bibinfo  {journal} {Phys. Rev. Lett.}\
  }\textbf {\bibinfo {volume} {109}},\ \bibinfo {pages} {165301} (\bibinfo
  {year} {2012})}\BibitemShut {NoStop}%
\bibitem [{\citenamefont {Liu}\ \emph {et~al.}(2010)\citenamefont {Liu},
  \citenamefont {Qi}, \citenamefont {Zhang}, \citenamefont {Dai}, \citenamefont
  {Fang},\ and\ \citenamefont {Zhang}}]{Liu2010}%
  \BibitemOpen
  \bibfield  {author} {\bibinfo {author} {\bibfnamefont {C.-X.}\ \bibnamefont
  {Liu}}, \bibinfo {author} {\bibfnamefont {X.-L.}\ \bibnamefont {Qi}},
  \bibinfo {author} {\bibfnamefont {H.~J.}\ \bibnamefont {Zhang}}, \bibinfo
  {author} {\bibfnamefont {X.}~\bibnamefont {Dai}}, \bibinfo {author}
  {\bibfnamefont {Z.}~\bibnamefont {Fang}}, \ and\ \bibinfo {author}
  {\bibfnamefont {S.-C.}\ \bibnamefont {Zhang}},\ }\href@noop {} {\bibfield
  {journal} {\bibinfo  {journal} {Phys. \ Rev. \ B}\ }\textbf {\bibinfo
  {volume} {82}},\ \bibinfo {pages} {045122} (\bibinfo {year}
  {2010})}\BibitemShut {NoStop}%
\bibitem [{\citenamefont {Qi}\ \emph {et~al.}(2008)\citenamefont {Qi},
  \citenamefont {Hughes},\ and\ \citenamefont {Zhang}}]{Qi2008}%
  \BibitemOpen
  \bibfield  {author} {\bibinfo {author} {\bibfnamefont {X.-L.}\ \bibnamefont
  {Qi}}, \bibinfo {author} {\bibfnamefont {T.~L.}\ \bibnamefont {Hughes}}, \
  and\ \bibinfo {author} {\bibfnamefont {S.-C.}\ \bibnamefont {Zhang}},\
  }\href@noop {} {\bibfield  {journal} {\bibinfo  {journal} {Phys. \ Rev.\ B.}\
  }\textbf {\bibinfo {volume} {78}},\ \bibinfo {pages} {195424} (\bibinfo
  {year} {2008})}\BibitemShut {NoStop}%
\bibitem [{\citenamefont {Tanaka}\ \emph {et~al.}(2009)\citenamefont {Tanaka},
  \citenamefont {Yokoyama},\ and\ \citenamefont {Nagaosa}}]{Tanaka2009}%
  \BibitemOpen
  \bibfield  {author} {\bibinfo {author} {\bibfnamefont {Y.}~\bibnamefont
  {Tanaka}}, \bibinfo {author} {\bibfnamefont {T.}~\bibnamefont {Yokoyama}}, \
  and\ \bibinfo {author} {\bibfnamefont {N.}~\bibnamefont {Nagaosa}},\
  }\href@noop {} {\bibfield  {journal} {\bibinfo  {journal} {Phys. \ Rev. \
  Lett.}\ }\textbf {\bibinfo {volume} {103}},\ \bibinfo {pages} {107002}
  (\bibinfo {year} {2009})}\BibitemShut {NoStop}%
\bibitem [{\citenamefont {Liu}\ \emph {et~al.}(2009)\citenamefont {Liu},
  \citenamefont {Liu}, \citenamefont {Xu}, \citenamefont {Qi}, ,\ and\
  \citenamefont {Zhang}}]{Liu2009}%
  \BibitemOpen
  \bibfield  {author} {\bibinfo {author} {\bibfnamefont {Q.}~\bibnamefont
  {Liu}}, \bibinfo {author} {\bibfnamefont {C.-X.}\ \bibnamefont {Liu}},
  \bibinfo {author} {\bibfnamefont {C.}~\bibnamefont {Xu}}, \bibinfo {author}
  {\bibfnamefont {X.-L.}\ \bibnamefont {Qi}}, , \ and\ \bibinfo {author}
  {\bibfnamefont {S.-C.}\ \bibnamefont {Zhang}},\ }\href@noop {} {\bibfield
  {journal} {\bibinfo  {journal} {Phys. \ Rev.\ Lett.}\ }\textbf {\bibinfo
  {volume} {102}},\ \bibinfo {pages} {156603} (\bibinfo {year}
  {2009})}\BibitemShut {NoStop}%
\bibitem [{\citenamefont {Wray}\ \emph {et~al.}(2010)\citenamefont {Wray},
  \citenamefont {Xu}, \citenamefont {Xia}, \citenamefont {Hsieh}, \citenamefont
  {Fedorov}, \citenamefont {Hor}, \citenamefont {Cava}, \citenamefont {Bansil},
  \citenamefont {Lin},\ and\ \citenamefont {Hasan}}]{Wray2010}%
  \BibitemOpen
  \bibfield  {author} {\bibinfo {author} {\bibfnamefont {L.~A.}\ \bibnamefont
  {Wray}}, \bibinfo {author} {\bibfnamefont {S.-Y.}\ \bibnamefont {Xu}},
  \bibinfo {author} {\bibfnamefont {Y.}~\bibnamefont {Xia}}, \bibinfo {author}
  {\bibfnamefont {D.}~\bibnamefont {Hsieh}}, \bibinfo {author} {\bibfnamefont
  {A.~V.}\ \bibnamefont {Fedorov}}, \bibinfo {author} {\bibfnamefont {Y.~S.}\
  \bibnamefont {Hor}}, \bibinfo {author} {\bibfnamefont {R.~J.}\ \bibnamefont
  {Cava}}, \bibinfo {author} {\bibfnamefont {A.}~\bibnamefont {Bansil}},
  \bibinfo {author} {\bibfnamefont {H.}~\bibnamefont {Lin}}, \ and\ \bibinfo
  {author} {\bibfnamefont {M.~Z.}\ \bibnamefont {Hasan}},\ }\href@noop {}
  {\bibfield  {journal} {\bibinfo  {journal} {Nat. \ Phys.}\ }\textbf {\bibinfo
  {volume} {7}},\ \bibinfo {pages} {32} (\bibinfo {year} {2010})}\BibitemShut
  {NoStop}%
\bibitem [{\citenamefont {Habe}\ and\ \citenamefont {Asano}(2012)}]{Habe2012}%
  \BibitemOpen
  \bibfield  {author} {\bibinfo {author} {\bibfnamefont {T.}~\bibnamefont
  {Habe}}\ and\ \bibinfo {author} {\bibfnamefont {Y.}~\bibnamefont {Asano}},\
  }\href@noop {} {\bibfield  {journal} {\bibinfo  {journal} {Phys. \ Rev.\ B.}\
  }\textbf {\bibinfo {volume} {85}},\ \bibinfo {pages} {195325} (\bibinfo
  {year} {2012})}\BibitemShut {NoStop}%
\bibitem [{\citenamefont {Murakami}(2007)}]{Murakami2007}%
  \BibitemOpen
  \bibfield  {author} {\bibinfo {author} {\bibfnamefont {S.}~\bibnamefont
  {Murakami}},\ }\href@noop {} {\bibfield  {journal} {\bibinfo  {journal} {New.
  J. Phys.}\ }\textbf {\bibinfo {volume} {9}},\ \bibinfo {pages} {356}
  (\bibinfo {year} {2007})}\BibitemShut {NoStop}%
\bibitem [{\citenamefont {Wan}\ \emph {et~al.}(2011)\citenamefont {Wan},
  \citenamefont {Turner}, \citenamefont {Vishwanath},\ and\ \citenamefont
  {Savrasov}}]{Wan2011}%
  \BibitemOpen
  \bibfield  {author} {\bibinfo {author} {\bibfnamefont {X.}~\bibnamefont
  {Wan}}, \bibinfo {author} {\bibfnamefont {A.~M.}\ \bibnamefont {Turner}},
  \bibinfo {author} {\bibfnamefont {A.}~\bibnamefont {Vishwanath}}, \ and\
  \bibinfo {author} {\bibfnamefont {S.~Y.}\ \bibnamefont {Savrasov}},\
  }\href@noop {} {\bibfield  {journal} {\bibinfo  {journal} {Phys. Rev. B}\
  }\textbf {\bibinfo {volume} {83}},\ \bibinfo {pages} {205101} (\bibinfo
  {year} {2011})}\BibitemShut {NoStop}%
\bibitem [{\citenamefont {Yang}\ \emph {et~al.}(2011)\citenamefont {Yang},
  \citenamefont {Lu},\ and\ \citenamefont {Ran}}]{Yang2011}%
  \BibitemOpen
  \bibfield  {author} {\bibinfo {author} {\bibfnamefont {K.-Y.}\ \bibnamefont
  {Yang}}, \bibinfo {author} {\bibfnamefont {Y.-M.}\ \bibnamefont {Lu}}, \ and\
  \bibinfo {author} {\bibfnamefont {Y.}~\bibnamefont {Ran}},\ }\href@noop {}
  {\bibfield  {journal} {\bibinfo  {journal} {Phys. Rev. B}\ }\textbf {\bibinfo
  {volume} {84}},\ \bibinfo {pages} {075129} (\bibinfo {year}
  {2011})}\BibitemShut {NoStop}%
\bibitem [{\citenamefont {Burkov}\ and\ \citenamefont
  {Balents}(2011)}]{Burkov2011}%
  \BibitemOpen
  \bibfield  {author} {\bibinfo {author} {\bibfnamefont {A.~A.}\ \bibnamefont
  {Burkov}}\ and\ \bibinfo {author} {\bibfnamefont {L.}~\bibnamefont
  {Balents}},\ }\href@noop {} {\bibfield  {journal} {\bibinfo  {journal} {Phys.
  Rev. Lett.}\ }\textbf {\bibinfo {volume} {107}},\ \bibinfo {pages} {127205}
  (\bibinfo {year} {2011})}\BibitemShut {NoStop}%
\bibitem [{\citenamefont {Liu}\ \emph {et~al.}(2013)\citenamefont {Liu},
  \citenamefont {Ye},\ and\ \citenamefont {Qi}}]{Liu2013}%
  \BibitemOpen
  \bibfield  {author} {\bibinfo {author} {\bibfnamefont {C.-X.}\ \bibnamefont
  {Liu}}, \bibinfo {author} {\bibfnamefont {P.}~\bibnamefont {Ye}}, \ and\
  \bibinfo {author} {\bibfnamefont {X.-L.}\ \bibnamefont {Qi}},\ }\href@noop {}
  {\bibfield  {journal} {\bibinfo  {journal} {Phys. Rev. B}\ }\textbf {\bibinfo
  {volume} {87}},\ \bibinfo {pages} {235306} (\bibinfo {year}
  {2013})}\BibitemShut {NoStop}%
\end{thebibliography}%

\end{document}